\hfuzz 2pt

\font\titlefont=cmbx10 scaled
\magstep1\magnification=\magstep1\null\vskip 1.5cm

\centerline{\titlefont DISSIPATIVE NEUTRINO OSCILLATIONS} 
\medskip
\centerline{\titlefont IN RANDOMLY FLUCTUATING MATTER}
\vskip 2.5cm

\centerline{\bf F. Benatti}\smallskip

\centerline{Dipartimento di Fisica Teorica, Universit\`a di Trieste}
\centerline{Strada Costiera 11, 34014 Trieste, Italy}
\centerline{and}
\centerline{Istituto Nazionale di Fisica Nucleare, Sezione di Trieste}
\vskip 1cm

\centerline{\bf R. Floreanini}
\smallskip

\centerline{Istituto Nazionale di Fisica Nucleare, Sezione di Trieste}
\centerline{Dipartimento di Fisica Teorica, Universit\`a di Trieste}
\centerline{Strada Costiera 11, 34014 Trieste, Italy}
\vskip 2.5cm

\centerline{\bf Abstract}\smallskip
\midinsert\narrower\narrower
\noindent
The generalized dynamics describing the propagation of neutrinos
in randomly fluctuating media is analyzed: it takes into
account matter-induced, decoherence phenomena that go beyond the
standard MSW effect. A widely adopted density fluctuation pattern
is found to be physically untenable: a more general model needs
to be instead considered, leading to flavor changing effective neutrino-matter
interactions. They induce new, dissipative effects that modify the 
neutrino oscillation pattern in a way
amenable to a direct experimental analysis.
\endinsert
\bigskip
\vfill\eject

\noindent
{\bf 1. INTRODUCTION} 
\medskip

When a neutrino propagates in a constant distribution of matter, coherent
forward scattering phenomena can affect its time evolution.
Despite the smallness of the cross-section induced by the neutrino
interaction with the medium, these matter effects can significantly modify
the oscillation pattern, through the so-called MSW mechanism.\hbox{[1, 2]}

However, forward scattering phenomena are just the simplest matter induced
effects that can occur to a neutrino when the medium is allowed to fluctuate.
In this case, the neutrino can be viewed as an open system, {\it i.e.}
a subsystem immersed in an external environment (the medium);[3-7] its time
evolution, obtained from the total neutrino+matter dynamics by
eliminating ({\it i.e.} integrating over) the matter degrees of freedom,
is no longer unitary: it takes into account possible exchanges of 
entropy and energy between the neutrino and the fluctuating medium.

In many physical situations, one can safely ignore the details of
the matter dynamics and use an effective description of the medium as
a classical, random external field. Quite in general, any environment
can be modeled in this way, provided the characteristic decay time
of the associated correlations is sufficiently small with respect to the
typical evolution time of the subsystem. In the case of
relativistic neutrinos, this time scale can be roughly identified with the
vacuum oscillation length: we shall therefore consider media that fluctuate on
time scales shorter than this. It has been recently pointed out that
the interior of the sun could indeed satisfy such a condition,[8] as
likely as the earth mantel. Thus, a neutrino created in the sun or moving through the
earth would effectively see a random fluctuating distribution of scattering
centers and therefore be subjected to stochastic, incoherent interaction
with the medium.
In this situation, correlations in the medium play a fundamental role:
they are responsible for the generation of new matter effects, beyond the
MSW ones, leading to irreversibility and loss of quantum coherence.

The effects of fluctuating matter on neutrino propagation have been
first discussed in [9-11], and recently reconsidered in [12-14]. However,
all these analysis deal with a simple density fluctuation pattern,
naturally suggested by the standard MSW treatment. Further, these fluctuations
are assumed to be exactly $\delta$-correlated; this is a highly
idealized description of the environment, that {\it e.g.}
for heat baths can be attained only in the limit 
of infinite temperature.

Instead, in the following a more realistic exponentially damped form for
the correlation functions in the medium will be adopted.
Limiting for simplicity the discussion to the analysis of the oscillations
of two species of neutrinos, we shall see that the effects induced
by matter fluctuations can be fully described in terms of a limited number of
phenomenological parameters. They affect the oscillation pattern
in a very distinctive way, that is amenable to a direct experimental
study.

On the other hand, when the simplified density fluctuation hypothesis
considered in [9-14] is adopted, a single constant is sufficient to
parametrize the new matter effects. However, this approximation appears
physically untenable, since by adopting it certain 
transition probabilities take unacceptable
negative values; this serious inconsistency can be cured only by allowing
more general matter fluctuations, pointing towards the presence
of flavor changing neutrino-matter interactions.%
\footnote{$^\dagger$}{Although within
a mean-field (MSW) approach, this possibility 
has recently been reconsidered in [15-17]
and found compatible with present experimental data.}

As a final remark, it is interesting to point out that the dissipative effects
induced by a randomly fluctuating medium on neutrino oscillations
involve in general the $CP$-violating phase that is present in the
mixing matrix for Majorana neutrinos. Therefore, contrary to the
vacuum case, matter oscillation experiments can provide, at least in
principle, a way to distinguish between Dirac and Majorana neutrinos.

\vskip 1cm

\noindent
{\bf 2. MASTER EQUATION}
\medskip

In discussing the mixing of two neutrino species, we shall adopt
the familiar effective description in terms of a two-dimensional
Hilbert space;[18-22] the flavor states, that we shall conventionally call
$|\nu_e\rangle$ and $|\nu_\mu\rangle$, will be chosen as basis states.
With respect to this basis, the physical neutrino states
are then represented by density matrices $R$, {\it i.e.} by
hermitian $2\times2$ matrices, with non-negative eigenvalues and unit trace.
Their time evolution equation can be cast in a standard
Liouville -- von Neumann form:[23, 24]
$$
{\partial R(t)\over \partial t}= -i\big[ H_0, R(t)\big]+ L_t\big[R(t)\big]\ .
\eqno(2.1)
$$
The first piece on the r.h.s. describes the propagation of the neutrinos 
in vacuum; in the chosen basis, the effective hamiltonian $H_0$
takes the standard form:
$$
H_0=\omega\, \vec n\cdot\vec\sigma\ ,
\eqno(2.2)
$$
where $\omega=\Delta m^2/4E$, $\Delta m^2$ being the square mass difference
of the two mass eigenstates and $E$ the average neutrino energy,
while the unit vector $\vec n=(\sin2\theta, 0, -\cos2\theta)$ contains
the dependence on the mixing angle, $\vec\sigma=(\sigma_1,\sigma_2,\sigma_3)$
being the vector of Pauli matrices.
The additional contribution $L_t[R]$ takes into account the presence
of matter. As explained above, we shall consider the case of
a rapidly fluctuating medium, which can be
described by classical stochastic fields.  Its action on the travelling
neutrinos can then be expressed via the commutator with a time-dependent
hermitian matrix $V(t)$,
$$
L_t\big[R(t)\big]=-i\big[ V(t), R(t) \big]\ ,\qquad V(t)=\vec V(t)\cdot\vec\sigma\ ,
\eqno(2.3)
$$
whose components $V_1(t)$, $V_2(t)$, $V_3(t)$ form a real, stationary Gaussian
stochastic field $\vec V(t)$; they are assumed to have in general a nonzero 
constant mean and translationally invariant correlations:
$$
\widehat{W}_{ij}(t-s)\equiv\langle V_i(t)\, V_j(s)\rangle
-\langle V_i(t)\rangle\, \langle V_j(s)\rangle\ ,\quad i,j=1,2,3\ .
\eqno(2.4)
$$

Since the generalized hamiltonian $V(t)$ in (2.3) involves stochastic 
variables,
the density matrix $R(t)$, solution of the equation of
motion (2.1), is also stochastic. Instead, we are interested in the behaviour
of the reduced density matrix $\rho(t)\equiv\langle R(t)\rangle$
which is obtained by averaging over the noise; it is $\rho(t)$ that
describes the effective evolution of the neutrinos in the medium and allows
the computation of relevant transition probabilities.
By making the additional assumption that neutrinos
and noise be decoupled at $t=\,0$, so that the initial state is
$\rho(0)\equiv\langle R(0)\rangle=R(0)$, a condition very well satisfied 
in typical situations, an effective master equation for $\rho(t)$
can be derived by going to the interaction representation,
where we set:
$$
\widetilde{R}(t)=e^{it\, H_0}\ R(t)\ e^{-it\, H_0}\ ,\quad
\vec\sigma(t)=e^{it\, H_0}\ \vec\sigma\ e^{-it\, H_0}\ ,\quad
\widetilde{L}_t[\ \ ]\equiv-i\Big[\vec V(t)\cdot\vec\sigma(t),\ \ \Big]\ .
\eqno(2.5)
$$
By averaging $\widetilde{R}(t)$ over the noise, we get the reduced density 
matrix $\tilde{\rho}(t)\equiv\langle\widetilde{R}(t)\rangle$ 
in the interaction representation: it is convenient to operate on the 
standard series expansion of $\widetilde{R}(t)$, so that:
$$
\tilde\rho(t)= {\cal N}_t[\tilde\rho(0)]\equiv
\sum_{k=0}^\infty N_t^{(k)}[\tilde\rho(0)]
\eqno(2.6a)
$$
where the terms $N_t^{(k)}$ are explicitly given by
$$
N_t^{(k)}[\tilde\rho(0)]=
\int_0^t ds_1\int_0^{s_1}ds_2\cdots \int_0^{s_{k-1}}ds_k\
\langle \widetilde{L}_{s_1} \widetilde{L}_{s_2}\cdots 
\widetilde{L}_{s_k}\rangle[\tilde\rho(0)\,]\ ,
\eqno(2.6b)
$$
with $N_t^{(0)}=1$ the identity, $N_t^{(0)}[\tilde\rho(0)]=\tilde\rho(0)$.
The resulting series is a sum over multiple integrals of
correlators $\langle \widetilde{L}_{s_1} \widetilde{L}_{s_2}\cdots
\widetilde{L}_{s_k}\rangle$, that is of averages
over the noise of successive commutators with respect to the
stochastic operators $V(t)=\vec V(t)\cdot\vec\sigma$ at different times.
The density matrix $\widetilde{\rho}(0)$ is not averaged over due to the 
assumption on the initial state $\rho(0)=\widetilde{\rho}(0)$.

In order to arrive at a more manageable time-evolution, we use a 
techique [23], of which we give a brief account below, that leads to a 
so-called convolution-less master equation and
is based on the hypothesis of weak coupling between system and stochastic 
environment.
The first step is to write the formal inverse of the map ${\cal N}_t$ in $(2.6a)$, 
{\it i.e.} $\tilde{\rho}(0)={\cal N}^{-1}_t[\tilde{\rho}(t)]$,
so that:
$$
{\cal N}^{-1}_t=\Bigl(1+\,N_t^{(1)}\,+\, N_t^{(2)}+\cdots\Bigr)^{-1}=
1\,-\,N_t^{(1)}\,-\,N_t^{(2)}-(N_t^{(1)})^2-\cdots\ ,
\eqno(2.7)
$$
where only terms containing up to two-point correlation functions have 
been indicated. Further, denoting 
with $\dot{N}^{(k)}_t$ the time-derivative of $N^{(k)}_t$, it follows that
the reduced density matrix in the interaction representation satisfies the
equation of motion
$$
{\partial \tilde{\rho}(t)\over\partial t}
={\partial{\cal N}_t\over\partial t}[\tilde{\rho}(0)]
={\partial{\cal N}_t\over\partial t}
\, {\cal N}_t^{-1}[\tilde\rho(t)]=
\Big\{\dot{N}_t^{(1)}\,+\,\big(\dot{N}_t^{(2)}
-\dot{N}_t^{(1)}\, N_t^{(1)}\big)+\ldots\Big\}[\tilde\rho(t)]\ .
\eqno(2.8)
$$

Since the interaction of the travelling neutrinos with the medium is weak,
one can focus the attention on the dominant terms of the previous expansion, 
neglecting all contributions higher than the second-order ones.
Further, since the characteristic decay time of correlations in the medium
is by assumption much smaller than the typical time scale of the system,
the memory effects implicit in (2.8) should not be physically relevant
and the use of the Markovian approximation justified. This is implemented 
in practice by extending to infinity the upper limit of the integrals
appearing in $\dot{N}^{(2)}$ and $N^{(1)}$.[3-5]

By returning to the Schr\"odinger representation, one finally obtains [24]
$$
{\partial\rho(t)\over\partial t}=-i\big[ H,\, \rho(t)\big]
+L\big[\rho(t)\big]\ , 
\eqno(2.9a)
$$
where
$$
\eqalignno{
&H=H_0 + H_1 + H_2\equiv\vec\Omega\cdot\vec\sigma\ , &(2.9b)\cr
&L[\rho]={1\over2}\sum_{i,j=1}^3 {\cal C}_{ij}
\Big[2\sigma_i \rho\, \sigma_j
- \{\sigma_j\sigma_i\, ,\, \rho\}\Big]\ . &(2.9c)
}
$$
The effective hamiltonian in matter, $H$, differs from the one in vacuum,
$H_0$, by first order terms (coming from the piece $\dot{N}^{(1)}$
in (2.8)) depending on the noise mean values:
$$
H_1= \langle \vec V(t)\rangle\cdot \vec\sigma\ ,
\eqno(2.10)
$$
and by second-order contributions (coming from the second-order terms 
in (2.8)),
$$
H_2=\sum_{i,j,k=1}^3\,\epsilon_{ijk}\ C_{ij}\, \sigma_k\ ,
\eqno(2.11)
$$
involving the noise correlations (2.4) through the 
time-independent combinations:
$$
C_{ij}=\sum_{k=1}^3\int_0^\infty dt\ \widehat{W}_{ik}(t)\ U_{kj}(-t)\ ,
\eqno(2.12)
$$
where the $3\times3$ orthogonal matrix $U(t)$ is defined by the following
transformation rule:
$e^{it\, H_0}\ \sigma_i\ e^{-it\, H_0}=\sum_{j=1}^3 U_{ij}(t)\ \sigma_j$.
On the other hand, the contribution $L[\rho\,]$ in $(2.9c)$
is a time-independent, trace-preserving linear map
involving the symmetric coefficient matrix ${\cal C}_{ij}\equiv C_{ij}+C_{ji}$.
It introduces irreversibility, inducing
in general dissipation and loss of quantum coherence.
Altogether, equation (2.9) generates a semigroup of linear maps,
${\mit\Gamma}_t:\ \rho(0)\mapsto\rho(t)\equiv{\mit\Gamma}_t[\rho(0)]$,
for which composition is defined only forward in time:
${\mit\Gamma}_t\circ{\mit\Gamma}_s={\mit\Gamma}_{t+s}$, with $t,s\geq0$;
this is a very general physical requirement that should be
satisfied by all Markovian open system dynamics.%
\footnote{$^\dagger$}{Notice that the procedure of averaging transition probabilities over
random matter profiles as performed in [25] is not compatible
with this basic evolution law, and therefore it is a pure phenomenological
device, hardly amenable to a rigorous theoretical treatment.}
The set of maps ${\mit\Gamma}_t$
is usually referred to as a quantum dynamical semigroup.\hbox{[3-7]}

The typical observable that is accessible to the experiments is the
probability ${\cal P}_{\nu_e\to\nu_\mu}(t)$ for having a transition to
a neutrino of type $\nu_\mu$ at time $t$, assuming that the neutrino 
has been generated as $\nu_e$ at $t=\,0$. In the language of density
matrices, it is given by:
$$
{\cal P}_{\nu_e\to\nu_\mu}(t)\equiv 
{\rm Tr}\big[\rho_{\nu_e}(t)\ \rho_{\nu_\mu}\big]\ ,
\eqno(2.13)
$$
where $\rho_{\nu_e}(t)$ is the solution of (2.9) with the
initial condition given by the matrix
$\rho_{\nu_e}(0)=\rho_{\nu_e}\equiv |\nu_e\rangle\langle\nu_e |$,
while $\rho_{\nu_\mu}=1-\rho_{\nu_e}$. By expanding the neutrino
density matrix in terms of the Pauli matrices and the identity $\sigma_0$,
$\rho=\big[\sigma_0+\vec\rho\cdot\vec\sigma\big]/2$,
the linear equation $(2.9a)$ reduces to a diffusion equation
for the components $\rho_1$, $\rho_2$, $\rho_3$ of the vector
$\vec\rho$:
$$
{\partial \vec\rho(t)\over\partial t}=-2{\cal H}\, \vec\rho(t)\ ;
\eqno(2.14)
$$
the entries of $3\times3$ matrix $\cal H$ can be expressed
in terms of the coefficients $\Omega_i$ and ${\cal C}_{ij}$
appearing in the hamiltonian and noise contribution in $(2.9b)$, $(2.9c)$:[26]
$$
{\cal H}=\left[
\matrix{a&b+\Omega_3&c-\Omega_2\cr     
b-\Omega_3&\alpha&\beta+\Omega_1\cr                                     
c+\Omega_2&\beta-\Omega_1&\gamma\cr}
\right]\ ,
\eqno(2.15)
$$
with $a={\cal C}_{22}+{\cal C}_{33}$, $\alpha={\cal C}_{11}+{\cal C}_{33}$,
$\gamma={\cal C}_{11}+{\cal C}_{22}$, $b=-{\cal C}_{12}$,
$c=-{\cal C}_{13}$, $\beta=-{\cal C}_{23}$. 
The solution of (2.14) involves the exponentiation of the matrix $\cal H$,
$$
\vec\rho(t)={\cal M}(t)\ \vec\rho(0)\ ,\qquad {\cal M}(t)=e^{-2{\cal H} t}\ ,
\eqno(2.16)
$$
so that the transition probability in (2.13) can be rewritten as
$$
{\cal P}_{\nu_e\to\nu_\mu}(t)=
{1\over2}\Big[1+\sum_{i,j=1}^3\rho^i_{\nu_e}\rho^j_{\nu_\mu}\ {\cal M}_{ij}(t)\Big]
={1\over2}\Big[1-{\cal M}_{33}(t)\Big]\ .
\eqno(2.17)
$$
Indeed, taking the standard form of the Pauli matrices with respect to the 
orthonormal basis $\vert\nu_e\rangle=\pmatrix{1\cr 0}$ and  
$\vert\nu_\mu\rangle=\pmatrix{0\cr 1}$, then
$\displaystyle \rho_{\nu_e}={1+\sigma_3\over2}$ and 
$\displaystyle \rho_{\nu_\mu}={1-\sigma_3\over2}$.

When correlations in the medium are negligible,
{\it i.e.} the combination in (2.4) are vanishingly small, 
equation (2.9) describes standard (MSW) matter effects, for
the presence of matter is signaled
solely by the shift $H_1$ in the effective hamiltonian.
In this case, the neutrino-medium interaction is dominated
by coherent forward scattering, and, in absence of flavor changing effects,
the stochastic vector field in (2.3) results oriented along the third axis,
whence $H_1=A\sigma_3$, where $A\equiv\langle V_3(t)\rangle=G_F n_e/\sqrt{2}$
gives the extra energy contribution that electron neutrinos receive
when travelling in ordinary matter
($G_F$ is the Fermi constant, while $n_e$ represents the electron
number density in the medium). 
As a consequence, 
the transition probability in (2.17) can be expressed in terms
of a modified frequency $\omega_M$ and mixing angle $\theta_M$
in matter,
$$
\omega_M=\omega \Big[\sin^22\theta +(1-A/A_R)^2 \cos^2 2\theta\Big]^{1/2}
\ ,\qquad
\sin 2\theta_M={\omega\over\omega_M}\sin 2\theta\ ,
\eqno(2.18)
$$
$A_R=\omega \cos2\theta$ being the value of $A$ at resonance.
In fact, the assumption of negligible correlations amounts to considering
in equation (2.14) a matrix $\cal H$ of the form
$$
{\cal H}=\pmatrix{0&\Omega_3&0\cr-\Omega_3&0&\Omega_1\cr0&-\Omega_1&0}\ ,\quad
\Omega_1=\omega\sin(2\theta)\ ,\ \Omega_3=A-\omega\cos(2\theta)\ .
\eqno(2.19)
$$ 
This matrix can be easily exponentiated as in (2.16),
$$
{\cal M}(t)=\pmatrix{{\Omega^2_1+\Omega_3^2\cos(2\omega_M t)\over\omega_M^2}&
-{\Omega_3\over\Omega_1}\sin(2\omega_M t)&
{\Omega_1\Omega_3\over\omega_M^2}(1-\cos(2\omega_M t))\cr
-{\Omega_3\over\Omega_1}\sin(2\omega_M t)&
\cos(2\omega_M t)&
-{\Omega_1\over\Omega_1}\sin(2\omega_M t)\cr
{\Omega_1\Omega_3\over\omega_M^2}(1-\cos(2\omega_M t))&
-{\Omega_1\over\Omega_1}\sin(2\omega_M t)&
{\Omega^2_3+\Omega_1^2\cos(2\omega_M t)\over\omega_M^2}
}\ ,
\eqno(2.20)
$$
whence the explicit form of the element ${\cal M}_{33}(t)$ yields
the familiar expression:
$$
{\cal P}_{\nu_e\to\nu_\mu}(t)=\sin^2 2\theta_M\ \sin^2 \omega_M t\ .
\eqno(2.21)
$$

The situation can significantly change for neutrinos immersed
in a fluctuating medium; while travelling in it, they encounter
matter fluctuations, whose correlations $\widehat W_{ij}(t-s)$ determine
the dissipative contribution in $(2.9c)$. 
In a typical bath at finite temperature, the correlation functions 
assume an exponentially damped form; therefore, one can generically
write:
$$
\widehat{W}_{ij}(t-s)=W_{ij}\ e^{-\lambda_{ij} |t-s|}\ ,
\eqno(2.22)
$$
with $W_{ij}$ and $\lambda_{ij}$ time-independent, real coefficients, 
with $\lambda_{ij}\geq0$. Further, as discussed before, the
stochastic medium fluctuates on time intervals much shorter than
the typical neutrino ``free'' evolution time scale $1/\omega$,
so that the decay parameters $\lambda_{ij}$ must be much larger
than the vacuum frequency $\omega$. This fact allows neglecting
all contributions higher than the first-order one in 
the ratio $\omega/\lambda_{ij}$
while evaluating the coefficients $C_{ij}$ in (2.12).
For generic correlations as in (2.22), these coefficients,
and therefore the entries of the matrix $\cal H$ in (2.15),
are all nonvanishing. However, the parameters
$a$, $b$, $c$, $\alpha$, $\beta$, $\gamma$ describing
matter decoherence effects are not all free:
as we shall see, physical consistency requires them 
to satisfy certain inequalities; in turn, these constraints
reflect some fundamental characteristics of the
matter-neutrino interactions.

We shall now discuss some interesting cases
of the master equation (2.9), corresponding to specific physical
realizations of the medium through which the neutrinos propagate.

\vskip 1cm

\noindent
{\bf 3. GENERALIZED MSW DYNAMICS}
\medskip

The simplest instance of a stochastic
medium corresponds to ordinary matter with density fluctuations,
where only the propagation of electron neutrinos is affected.
It generalizes the familiar MSW mean field treatment by 
adding to it decoherence effects.
In this case, the stochastic hamiltonian in (2.3) becomes
diagonal and, without loss of generality,
only the stochastic field $V_3(t)$ can be taken to be non-vanishing;
the neutrinos are still forward scattered by the medium,
although no longer in a coherent way.
This is situation discussed in [9-14], where however the density
fluctuations in the medium are taken to be exactly $\delta$-function
correlated. This is a highly idealized assumption, that can hardly
be reproduced in ordinary conditions. Instead, the much more realistic
exponential ansatz (2.22) will be used here, where the only nonvanishing
correlation strength and decay constant are $W_{33}\equiv W$ and
$\lambda_{33}\equiv\lambda$, respectively.

The noise contributions in (2.9) can be explicitly computed;
within our approximation,
one finds that only the entries ${\cal C}_{23}$ and ${\cal C}_{33}$
of the coefficient matrix in $(2.9c)$ are nonvanishing,
$$
a={\cal C}_{33}={2W\over\lambda}\ ,\qquad 
\beta=-{\cal C}_{23}={\omega W\over\lambda^2}\sin2\theta\ ,
\eqno(3.1)
$$
while the hamiltonian contribution $H_1$ is proportional to $\sigma_3$
(the standard MSW piece) and $H_2$ to $\sigma_1$:
$$
\Omega_1=\omega\Bigg(1+{W\over\lambda^2}\Bigg)\sin2\theta\ ,\quad
\Omega_2=\, 0\ ,\quad
\Omega_3=-\omega\Bigg(1-{A\over A_R}\Bigg)\cos2\theta\ .
\eqno(3.2)
$$
Surprisingly, the dynamics generated by (2.9), or equivalently (2.14), 
with these coefficients appears to be physically unacceptable.

As mentioned at the beginning, any density matrix must be a positive operator
({\it i.e.} its eigenvalues should be non-negative) in order to
represent a physical state: its eigenvalues have the physical meaning
of probabilities. Therefore, any time evolution needs to preserve this property,
otherwise an initial state would not be mapped to another state at
a later time. This is precisely what happens when the neutrino evolution
in the medium is modeled by (2.9) with dissipative parameters as in (3.1).
In fact, the probability ${\cal P}(t)$ for having a transition from
an initial neutrino state $\rho(0)$ to its orthogonal state
$\rho_\perp\equiv 1-\rho(0)$ at a later time $t$ is given by
the first equality in (2.17), with the substitutions
$\rho_{\nu_e}\to\rho(0)$, $\rho_{\nu_\mu}\to\rho_\perp$. 
Since ${\cal P}$ is initially zero,
its time derivative must be positive at $t=\,0$, otherwise
we would have physically unacceptable negative transition
probabilities as soon as $t>0$. A simple computation gives:
$\dot{\cal P}(0)=\sum_{i,j=1}^3\rho(0)_i\, {\cal H}_{ij}\, \rho(0)_j\geq0$,
and since this must be true for any initial state, physical
consistency requires the symmetric part of the matrix $\cal H$ in (2.15)
to be positive. One easily sees that this is impossible with
the assignment in (3.1).%
\footnote{$^\dagger$}{An example of the emergence of negative transition
probabilities is explicitly provided in the Appendix.}

In the case of a $\delta$-correlated medium,
the parameter $\beta$ identically vanishes and no inconsistencies arise;
however, as mentioned before, this choice is not supported by strong
physical motivations and appears just a
mathematically convenient simplification.
By naively relaxing the $\delta$-correlated assumption, one ends up
with the simple stochastic system discussed above,
which turns out to be seriously flawed. As a consequence,
modelling matter fluctuations only in terms
of electron density is physically untenable and indicates
that in order to consistently describe neutrino oscillations
in random matter more complex situations need to be analyzed,
involving a richer covariance structure than with
$\vec V(t)=(0,0,V_3(t))$.

Alternatively, instead of the random matter model one may question 
the approximations used in deriving the master equation (2.9), and precisely
the weak coupling hypothesis and the markovian limit.
However, the first assumption appears rather well satisfied in the case
of the neutrinos, as they interact very weakly with matter,
while the markovian approximation is justified by the physically motivated 
choice
of rapidly decaying matter correlations: $\lambda\gg\omega$.
In reality, once a slightly generalized model of
random medium is adopted, the master equation (2.9)
results perfectly adequate to consistently treat decoherence phenomena
in neutrino matter oscillations.

\vskip 1cm

\noindent
{\bf 4. DIAGONAL CORRELATIONS} 
\medskip

When the components of the stochastic field $\vec V(t)$
are all nonvanishing, the noise hamiltonian in (2.3) is no longer diagonal:
in this case, while travelling in the medium,
all neutrino species undergo incoherent scatterings, in general involving
not exclusively the forward direction; this may happen only in presence
of flavor changing interactions.
However, as a minimal extension of the previously
treated case, we shall assume $V_1(t)$ and $V_2(t)$ to have zero mean,
so that the hamiltonian correction $H_1$ contains only the standard MSW 
contribution, and further take the correlation functions in (2.22) 
to be diagonal:
$$
\widehat{W}_{ij}(t-s)=W_i\ e^{-\lambda_i |t-s|}\, \delta_{ij}\ .
\eqno(4.1)
$$
In addition, for simplicity we shall consider situations for which
the ratios $W_i/\lambda_i$ are all equal to a common factor ${\cal W}>0$;
in this case, the parameters appearing in (2.15) take the form:
$$
a=\alpha=\gamma=4{\cal W}\ ,\quad 
b=2\omega {\cal W}\Bigg({1\over\lambda_1}-{1\over\lambda_2}\Bigg)\cos2\theta\ ,\quad
c=\,0\ ,\quad
\beta=2\omega {\cal W}\Bigg({1\over\lambda_3}-{1\over\lambda_2}\Bigg)\sin2\theta\ ,
\eqno(4.2a)
$$
$$
\Omega_1=\omega\Bigg[1+2{\cal W}\Bigg({1\over\lambda_2}+{1\over\lambda_3}\Bigg)
\Bigg]\sin2\theta\ ,\ \
\Omega_2=\,0\ ,\ \
\Omega_3=-\omega\Bigg[1-{A\over A_R}+2{\cal W}\Bigg({1\over\lambda_1}+{1\over\lambda_2}\Bigg)
\Bigg]\cos2\theta\ ,
\eqno(4.2b)
$$
and the master equation (2.9) can be exactly integrated. Notice that
the request of positivity of $\rho(t)$ for any $t\geq0$ now requires
$\alpha^2\geq b^2 +\beta^2$, condition that is always satisfied
by the original hypothesis of fast decaying matter
correlations: $\lambda_i\gg\omega$. Even more, this inequality
guarantees not only the positivity of the evolution generated by (2.9),
but actually a stronger attribute, that of ``complete positivity''.[3-5]
This property is crucial in assuring the consistency of any
generalized, dissipative dynamics in all possible physical conditions
and should always be imposed in place of simple positivity
to avoid possible inconsistencies in the treatment;[26]
it is reassuring that it emerges naturally from our simple
model of random matter, without the need of further assumptions.

The transition probability ${\cal P}_{\nu_e\to\nu_\mu}$ in (2.17)
can be explicitly computed and cast in the simple form:
$$
{\cal P}_{\nu_e\to\nu_\mu}(t)={1\over2}\Big(1-e^{-2\alpha t}\Big)
+e^{-2\alpha t}\,\sin^2 2\tilde\theta\, \sin^2\Omega t\ ,
\eqno(4.3)
$$
where $\Omega=[\Omega_1^2+ \Omega_3^2 -b^2-\beta^2]^{1/2}$ is the modified
oscillation frequency, while
$\sin^2 2\tilde\theta=(\Omega^2_1-\beta^2)/\Omega^2\leq 1$ defines a new
mixing angle (notice that the absence of the parameters $c$ and $\Omega_2$ 
is due to the assumptions that led to $(4.2a,b)$).
In comparison with the standard result in (2.21), one sees that
the presence of a random medium introduces exponential
damping terms and further modifies the neutrino effective masses and
mixing properties; a resonance enhancement is still present
for $A=A_R$, but its effectiveness is reduced by the damping factors.
This is even more dramatic at large times, where the decoherence 
effects dominate: the neutrino state $\rho$ is driven to
the totally mixed state $\sigma_0/2$ and the transition probability approaches
its asymptotic $1/2$ value.

These conclusions apply to neutrinos travelling in uniform random media. 
When this is not the case,
the neutrino total time evolution results from the composition
of arbitrarily many partial evolutions corresponding to media with uniform 
properties,
but in general of different thicknesses; then, the complete evolution matrix 
${\cal M}(t)$
as defined in (2.16) will be the result of the composition
of the corresponding ones pertaining to the various media (a simple example
is given in the Appendix).  
Nevertheless, for slowly varying conditions, this composition
can be well approximated by its adiabatic expression,
obtained by the instantaneous diagonalization of the 
now time-dependent matrix
$\cal H$ in (2.15) and the assumption that the neutrino states
evolve as one of its eigenstates.%
\footnote{$^\dagger$}{Possible hoppings among the instantaneous eigenstates
can also be easily included; for simplicity, we ignore them here.}
Within this approximation and neglecting fast oscillating terms,
the averaged transition probability 
can be cast in the following form:
$$
{\cal P}_{\nu_e\to\nu_\mu}(t)={1\over2}
\Bigg[1-{e^{-8{\cal W} t}\over{\cal R}}\,
\Bigg(1-{A\over A_R}+{4{\cal W}\over\lambda_2}\Bigg)\ 
\cos 2\theta \Bigg]\ ,
\eqno(4.4)
$$
with
$$
{\cal R}=\Bigg\{ \Bigg[1+{4{\cal W}\over\lambda_2}
\Bigg(1+{\lambda_2\over\lambda_3}+{4{\cal W}\over\lambda_3}\Bigg)\Bigg]\tan^2 2\theta
+\Bigg(1-{A\over A_R}\Bigg)\Bigg[ 1-{A\over A_R} +
{4{\cal W}\over\lambda_2}
\Bigg(1+{\lambda_2\over\lambda_1}+{4{\cal W}\over\lambda_1}\Bigg)\Bigg]
\Bigg\}^{1/2}\ .
\eqno(4.5)
$$

With respect to standard, familiar expressions, the action of the stochastic
medium is signaled by the presence in the second term of a modified weight and
a damping factor; these additional contributions depend on the ratios of the three
matter-correlations decay constants $\lambda_i$ and the
corresponding strength $\cal W$. Although in the weak-coupling regime
one expects $\cal W\ll\omega$,
the decay constant ratios need not be small. Therefore, the behaviour of (4.4)
as a function of the neutrino energy can sensibly differ from the 
one obtained in absence of decoherence effects (concrete examples
are shown in Fig.1). 

Of particular interest
is the application of (4.4) to the solar neutrino case, where
$\Delta m^2$ and $\theta$ can be taken to assume the
best fit values obtained in recent data analysis 
({\it e.g.} see [27, 28] and references therein);
thanks to the availability of a larger decoherence parameter space,
the electron surviving probability 
${\cal P}_{\nu_e\to\nu_e}\equiv 1-{\cal P}_{\nu_e\to\nu_\mu}$
is found to differ not only from the
standard, noiseless expression, but also from those obtained with
$\delta$-correlated fluctuating matter
as reported in [13, 14] (for a comparison, see Fig.1).
These results, together with the still present uncertainties in the 
fluctuating behaviour
of the solar matter, appear to open concrete possibilities for
an experimental study of matter induced effects in neutrino oscillations
that go beyond the standard MSW phenomenology.

\vskip 1cm

\noindent
{\bf 5. DISCUSSION}
\medskip

In the most general situation, the correlations in
the stochastic medium have the form (2.22) and thus all the entries of
the matrix $C_{ij}$ in (2.12) result nonvanishing; as a consequence,
all second order pieces in the effective hamiltonian $(2.9b)$
as well as in the dissipative part $(2.9c)$ will contribute
to the master equation $(2.9a)$. Further, the first order
mean field approximation in (2.10) will no longer be 
diagonal, taking into account the presence of possible flavor changing
interactions.[15-17] 

Nevertheless, even in this very general case,
the corresponding matrix $\cal H$ in (2.15) can not result
totally generic: as already pointed out,
the positivity of the evolved state $\rho(t)$ must be preserved
under all circumstances; this is guaranteed by the
mentioned condition of complete positivity of the evolution 
generated by (2.9). This property requires
the positivity of the matrix ${\cal C}_{ij}$
in $(2.9c)$ and as a consequence imposes certain inequalities
among the dissipative parameters in (2.15)
(see [29, 30] for explicit expressions). These conditions
are certainly of help in restricting the parameter space
needed to describe a totally generic random medium.

Even with these constraints, no simple, exact analytic expressions for 
the transition probability 
${\cal P}_{\nu_e\to\nu_\mu}(t)$ in (2.17) can in general be given.
However, as discussed before,
second-order matter contributions to $\cal H$
are small with respect to the vacuum frequency $\omega$;
therefore, in solving (2.14) one can integrate the hamiltonian
dynamics exactly,
while treating the dependence on
$a$, $b$, $c$, $\alpha$, $\beta$, $\gamma$
in perturbation theory.%
\footnote{$^\dagger$}{The effects of the hamiltonian corrections 
to the free motion are in general not small, in particular near resonance;
this is why no approximation is allowed in the evolution
generated by the effective hamiltonian $(2.9b)$.}
In this way manageable, approximate
expressions for the transition probabilities can be obtained.
Having now at disposal a larger parameter space, their
form involves multiple damping factors and oscillation phases,
showing possible larger deviations from the standard behaviour.
In addition, notice that in order to describe neutrino mixing 
in a generic random medium
two mixing angles, $\hat\theta$ and $\hat\varphi$, are in general needed:
they parametrize the components of the unit vector
$\Omega_i/\Omega\equiv\big(\cos\hat\varphi\sin2\hat\theta, 
\sin\hat\varphi\sin2\hat\theta, -\cos2\hat\theta\big)$, with
$\Omega=|\vec\Omega|$,
which identifies the effective hamiltonian 
$H=\vec\Omega\cdot\vec\sigma$ in $(2.9b)$.

Actually, in presence of Majorana neutrinos,
also in vacuum the most general mixing matrix
involves two angles, $\theta$ and $\varphi$, so that
the explicit expression of the free effective hamiltonian 
$H_0$ in terms of these angles is as for $H$ above.
Although for oscillations in vacuum involving only
two species of neutrinos the angle $\varphi$
disappears from all observables, this is no longer
true in presence of matter induced decoherence effects.
Indeed, one can directly check that the transition probabilities 
explicitly depend
on $\varphi$, unless the dissipative parameters
$a$, $b$, $c$, $\alpha$, $\beta$, $\gamma$ are all zero;
at least in principle, it is therefore possible
to distinguish between Dirac and Majorana neutrinos
by studying their oscillations in random matter.
The detailed analysis of such of dependence
is certainly beyond the scope of the present
investigation and thus, in order to keep the
treatment as simple as possible, 
in the previous discussions we have tacitly assumed the neutrinos travelling
in matter to be of Dirac type, setting
$\varphi=\,0$ from the beginning.

As a final remark, let us mention that master
equations of the type (2.9) 
generate the most general open system dynamics compatible
with a semigroup composition law and the requirement of complete positivity,
and as such can be applied to model in a physically consistent way
a wide range of phenomena.[3-7]
In particular, they have been recently used in order to describe dissipative 
effects induced at low energies by the dynamics of fundamental objects 
(strings and branes) at a very high scale, typically the Planck mass.[29, 30] 
These string induced decoherence effects may modify the pattern
of neutrino oscillations,
and in principle interfere with the phenomena described above.
Nevertheless, besides being very small, they
affect in equal manner all types of neutrinos, so that they can be
isolated from the matter-induced effects by analyzing data taken
in different experimental conditions.

\vskip 1.5cm

\noindent
{\bf APPENDIX}
\medskip

In order to show that negative probabilities arise
in experimental accessible observables once the naive model
of density fluctuating matter discussed in the text is adopted,
one needs to combine neutrino propagation in vacuum with 
that in the medium. Consider a neutrino, created as
$\nu_e$, that propagates for time $t'$ in vacuum, then enters
the random medium in which stays for a time $t$, and is finally
detected after having travelled again in vacuum for a further time $t''$.
The probability ${\cal P}_{\nu_e\to\nu_\mu}(\tau)$ 
of finding a neutrino of type $\nu_\mu$
at the final time $\tau=t'+t+t''$ can be expressed as in (2.17),
where the total transition matrix ${\cal M}(\tau)$
is now the product of three terms,
${\cal M}(\tau)={\cal M}_0(t')\cdot{\cal M}(t)\cdot{\cal M}_0(t'')$,
the middle representing the propagation in the medium
with parameters as in (3.1) and (3.2), while the outer two
the ``free'' motion in vacuum, generated by the hamiltonian (2.2).

When the vacuum evolution time $t'$ is chosen to be very short, such that
$\sin\omega t'=\beta/(2a\sin2\theta)=\omega/4\lambda$, the state 
of the neutrino entering the medium is 
$\rho_-=[\sigma_0-\vec\rho_-\cdot\vec\sigma]/2$,
where $\vec\rho_-=-{\cal M}_0(t') \vec\rho_{\nu_e}(0)$ 
coincides with the eigenvector of the dissipative part 
of $\cal H$ relative to its negative eigenvalue. 
Similarly, with the same choice also for $t''$,
one finds ${\cal M}_0(-t'') \vec\rho_{\nu_\mu}(0)=\vec\rho_-$, so that
when exiting the medium the neutrino is found in the state $\rho_+=1-\rho_-$,
orthogonal to $\rho_-$.
With these conditions, one has:
${\cal P}_{\nu_e\to\nu_\mu}(\tau)={\cal P}_{-\to +}(t)$,
and near resonance, one explicitly finds:
$$
{\cal P}_{\nu_e\to\nu_\mu}(\tau)={1\over 2}\Bigg[1-e^{- a t}
\Bigg(\cos 2\Omega t + D\ {\sin 2\Omega t\over\Omega}\Bigg)\Bigg]\ ,
$$
where $D=[a^2/4+\beta^2]^{1/2}$ and $\Omega=[\omega^2-D^2]^{1/2}$; 
this expression
indeed assumes unphysical negative values for sufficiently small times:
${\cal P}_{\nu_e\to\nu_\mu}(\tau)\simeq (a/2-D)t$.

\vfill\eject

\line{}
\centerline{\bf ACKNOWLEDGMENT}
\bigskip

We thank A. Yu. Smirnov for reading the manuscript and for very useful discussions.

\vskip 3.5cm

\centerline{\bf REFERENCES}
\vskip 1cm

\item{1.} L. Wolfenstein, Phys. Rev. D {\bf 17} (1978) 2369;
{\it ibid.} {\bf 20} (1979) 2634
\smallskip
\item{2.} S.P. Mikheyev and A.Yu. Smirnov, Sov. J. Nucl. Phys.
{\bf 42} (1985) 913; Nuovo Cim. {\bf 9C} (1986) 17
\smallskip
\item{3.} R. Alicki and K. Lendi, {\it Quantum Dynamical Semigroups and Applications}, Lect. Notes Phys. {\bf 286}, (Springer, Berlin, 1987)\smallskip\item{4.} V. Gorini, A. Frigerio, M. Verri, A. Kossakowski andE.C.G. Surdarshan, Rep. Math. Phys. {\bf 13} (1978) 149 \smallskip\item{5.} H. Spohn, Rev. Mod. Phys. {\bf 52} (1980) 569\smallskip
\item{6.} H.-P. Breuer and F. Petruccione, {\it The Theory of Open
Quantum Systems}, (Oxford University Press, Oxford, 2002)
\smallskip
\item{7.} {\it Dissipative Quantum Dynamics}, F. Benatti and R. Floreanini eds.,
Lect. Notes Phys. {\bf 612}, (Springer, Berlin, 2003)
\smallskip
\item{8.} C.P. Burgess, N.S. Dzhalilov, T.I. Rashba, V.B. Semikoz and J.W.F. Valle,
Mon. Not. Roy. Astron. Soc. {\bf 348} (2004) 609
\smallskip
\item{9.} F.N. Loreti and A.B. Balantekin, Phys. Rev. D {\bf 50} (1994) 4762
\smallskip
\item{10.} H. Nunokawa, A. Rossi, V.B. Semikoz and J.W.F. Valle,
Nucl. Phys. {\bf B472} (1996) 495
\smallskip
\item{11.} E. Torrente-Lujan, Phys. Rev. D {\bf 59} (1999) 073001
\smallskip
\item{12.} C.P. Burgess, N.S. Dzhalilov, M. Maltoni, T.I. Rashba, V.B. Semikoz,
M. Tortola and J.W.F. Valle, Astrphys. J. {\bf 588} (2003) L65
\smallskip
\item{13.} A.B. Balantekin and H. Y\"uksel, Phys. Rev. D {\bf 68} (2003)
013006
\smallskip
\item{14.} M.M. Guzzo, P.C. de Holanda and N. Reggiani,
Phys. Lett. {\bf B569} (2003) 45
\smallskip
\item{15.} A. Friedland, C. Lunardini and C. Pena-Garay,
Phys. Lett. {\bf B594} (2004) 347
\smallskip
\item{16.} M.M Guzzo, P.C. de Holanda and O.L.G. Peres,
Phys. Lett. {\bf B591} (2004) 1
\smallskip
\item{17.} K.M. Zurek, JHEP {\bf 04} (2004) 058
\smallskip
\item{18.} S.M. Bilenky and S.T. Petcov, Rev. Mod. Phys. 
{\bf 59} (1987) 671
\smallskip
\item{19.} T.K. Kuo and J. Pantaleone, Rev. Mod. Phys. {\bf 61} (1989) 937
\smallskip
\item{20.} C.W Kim and A. Pevsner, {\it Neutrinos in Physics and 
Astrophysics}, (Harwood Academic Press, 1993)
\smallskip
\item{21.} R.N. Mohapatra and P.B. Pal, {\it Massive Neutrinos in Physics
and Astrophysics}, 2nd ed., (World Scientific, Singapore, 1999)
\smallskip
\item{22.} M. Fukugita and T. Yanagida, {\it Physics of Neutrinos and 
Applications to Astrophysics}, (Springer, Berlin, 2003)
\smallskip
\item{23.} J. Budimir and J.L. Skinner, J. Stat. Phys. {\bf 49}
(1987) 1029
\smallskip\item{24.} F. Benatti, R. Floreanini and R. Romano, J. Phys. A
{\bf 35} (2002) 4955
\smallskip
\item{25.} F.N. Loreti, Y.ÐZ. Qian, G.M. Fuller and A.B. Balantekin, 
Phys. Rev. D {\bf 52} (1995) 6664
\smallskip
\item{26.} F. Benatti and R. Floreanini, Mod. Phys. Lett. {\bf A12} (1997) 1465;
Banach Center Publications, {\bf 43} (1998) 71;
Phys. Lett. {\bf B468} (1999) 287; 
Chaos, Solitons and Fractals {\bf 12} (2001) 2631
\smallskip
\item{27.} V. Barger, D. Marfatia and K. Whisnant, 
Int. J. Mod. Phys. {\bf E12} (2003) 569
\smallskip
\item{28.} R.D. McKeown and P. Vogel, Phys. Rep. {\bf 394} (2004) 315
\smallskip
\item{29.} F. Benatti and R. Floreanini, JHEP {\bf 02} (2000) 032
\smallskip
\item{30.} F. Benatti and R. Floreanini, Phys. Rev. D {\bf 64} (2001)
085015

\vfill\eject

\vskip 3cm

\input epsf
\centerline{
\epsfxsize=10cm
\epsfbox{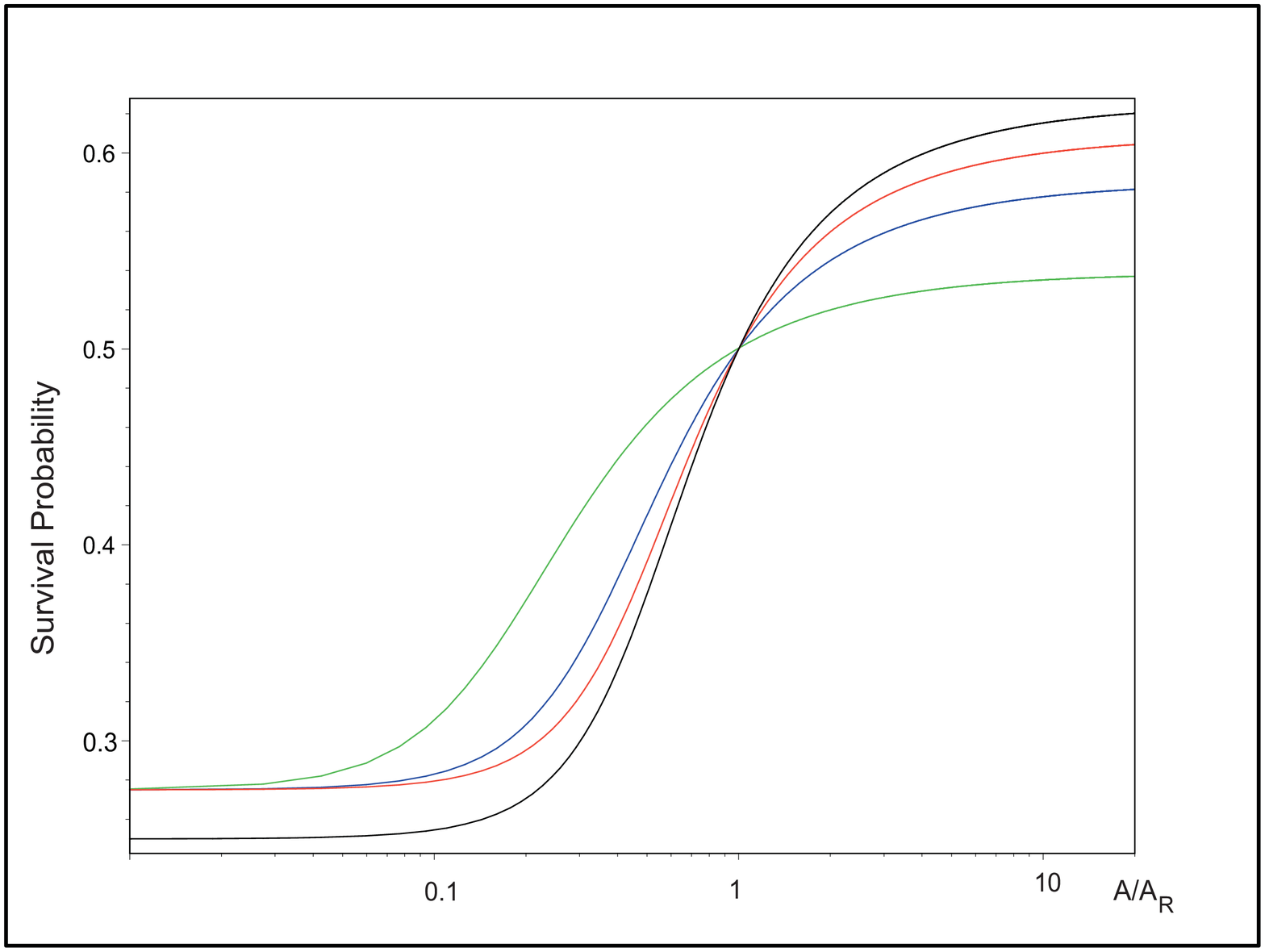}
}
\midinsert
\narrower\noindent
{\bf Figure 1.} Behaviour of electron neutrino mean survival
probability ${\cal P}_{\nu_e\to\nu_e}$ as a function of the neutrino energy 
(through the ratio $A/A_R$), for $\sin^2 2\theta\simeq 0.8$, 
density dominated matter fluctuations,
$\lambda_1,\ \lambda_2\gg\lambda_3$, and different correlation strengths,
${\cal W}/\lambda_2\simeq 10^{-4}\div 10^{-3}$. The lower starting (black) curve corresponds to
the case of noisless matter (standard MSW effect), while the remaining
(colored) ones show the effect of the stochastic fluctuations.
The initial gap among the group of curves is due to the presence
of the decoherence driven damping factor.

\endinsert

\bye